\documentclass{aastex631}

\begin{document}

\title{Mass and Color Dependence of the Hubble Spiral Sequence}

\author{Petra Mengistu}
\affiliation{Haverford College \\ 370 Lancaster Avenue \\Haverford, PA 19041, USA}

\author{Karen L. Masters}
\affiliation{Haverford College \\ 370 Lancaster Avenue \\Haverford, PA 19041, USA}

\begin{abstract}
In the classic Hubble spiral sequence, arm windiness correlates with bulge size; Sa type spiral galaxies with larger bulges also have the most tightly wound spirals. Exceptions to this have long been known, and in recent work using Galaxy Zoo morphologies no strong correlation was seen in a volume limited sample. In this Research Note, we explore the impact of galaxy mass and integrated color upon this correlation in the Galaxy Zoo sample, finding that bluer and lower mass spirals show the ``expected" correlation; however, it becomes slightly negative for redder and/or more massive spiral galaxies. 
\end{abstract}



\section{Introduction} \label{sec:intro}
Hubble’s classification scheme \citep{Hubble1926} classifies galaxies into two major categories: elliptical and spirals. Spirals are ordered in increasing order of spiral arm windiness, ``concentration" of the arms and dominance of the central bulge size. From this, it has been implicitly assumed that arm windiness positively correlates with galaxy bulge size and some work has measured this \citep[e.g.][]{Kennicutt1982}. The correlation motivated, in part, early models of static density wave spiral arms \citep[i.e.][]{LinShu1964} which predict tighter arms in more centrally concentrated galaxies. 
 However, it has long been known that the diversity of spiral galaxy morphologies is not captured by the Hubble sequence (e.g. see \citealt{Buta2013}), and current understanding of spiral arms has moved on from static density waves (e.g. see \citealt{SellwoodMasters2022}).  
Using a volume limited sample of galaxies selected from the Galaxy Zoo 2 (GZ2) survey, and defining an arm winding and bulge size score, \citet{Masters2019} found no significant correlation. 

In this Research Note, we seek to better understand how spiral arm winding and bulge size are related in sub-samples of spirals separated by global properties (namely mass and colour).

\section{Data and Sample Selection} \label{sec:style}

Our base sample is GZ2 \citep{Willett2013} -- 250,000 nearby galaxies with Sloan Digital Sky Survey (SDSS) Legacy imaging and spectroscopy \citep{2000AJ....120.1579Y}. 

We limit to $z<0.035$ and $\log(M_\star/M_\odot)>9.0$, for a volume-limited sample of $N=$21,899. We use debiased vote fractions, `p$_{\rm [primary response]}$', from GZ2 \citep{Willett2013} 
 
to select 7,272 spiral galaxies using $p_{\rm features}>0.430$; $p_{\rm notedge\_on}>0.715$, and $p_{\rm spiral}>0.5$. 

 We use a previously defined arm winding score, $w_{\rm avg}=0.5p_{\rm medium spiral} + 1.0p_{\rm tight spiral}$, and bulge prominence score, $b_{\rm avg}=0.2p_{\rm just noticeable bulge} + 0.8p_{\rm obvious bulge} + 1.0p_{\rm dominant bulge}$. These values have been previously defined and are discussed in \citet{Hart2017,Masters2019}. The choice of coefficients is arbitrary; these values produce scores ranging from 0--1 and were calibrated against other measures of pitch angle and bulge size in \citet{Hart2017}.

We make four subsets of spiral galaxies based on mass and global color as illustrated in Figure \ref{fig:subsets}. We use NUV-$r$ colours and stellar mass from the NASA Sloan Atlas \citep[NSA, ][namely ELPETRO$\_$MASS]{Blanton2005}. We cut at (NUV-$r$) = 4 and $\log(M_\star/M_\odot)= 10.3$ motivated by the typical colour and mass of the blue cloud--red sequence transition (e.g. \citealt{Chilingarian2012, Smethurst2022}). 

Overall, we have a total of 7,272 spiral galaxies divided into 5,999 low mass blue, 758 massive blue, 297 low mass red and 218 massive red spirals.

\section{Results} 
\begin{figure*}
\centering
	\includegraphics[width=18cm]{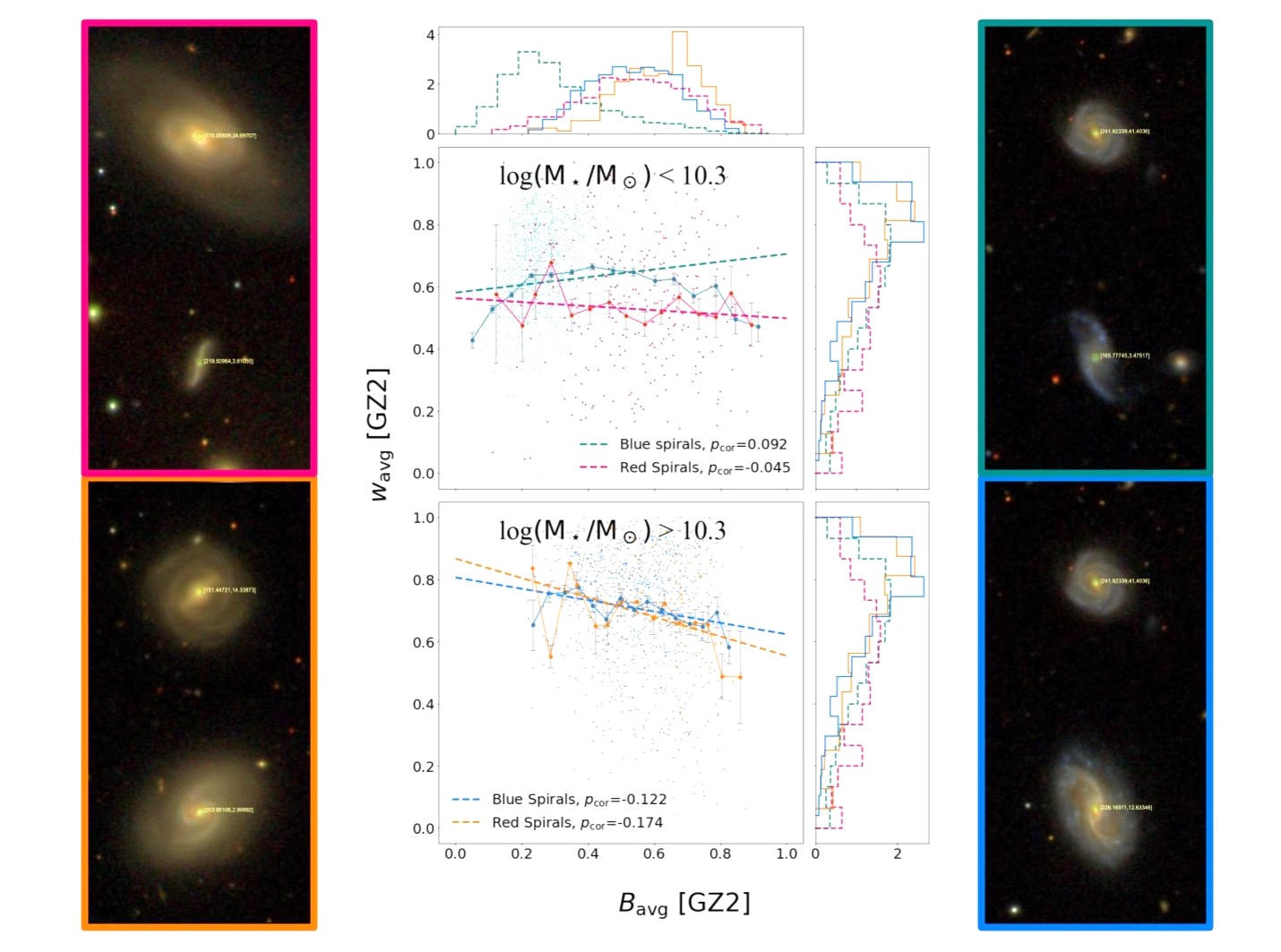}
    \caption{[Middle]: Two subplots showing winding score, w$_{\rm avg}$ against bulge size, b$_{\rm avg}$ for subsets of different mass and color spirals. The small points show the raw data, to which we have fitted a trend line; we also show fifteen equally spaced binned averages - error bars show standard error on the mean.  [Upper middle]: Low mass $\log(M_\star/M_\odot< 10.3)$ sample of galaxies divided into blue (N$_\mathrm{gal}$=5999), and red (N$_\mathrm{gal}$=297) spirals (dark cyan and pink data points and lines). [Lower middle]: High mass $\log(M_\star/M_\odot> 10.3)$ subset subdivided into blue (N$_\mathrm{gal}$=758), and red (N$_\mathrm{gal}$=218) spirals (blue and orange data points and lines). Histograms of bulge size and winding score are also shown for all subsets. [Side]: Example images of tightly and loosely wound spiral galaxies in each subset; blue spirals on the right, red at left, low mass at upper, and high mass lower.}
    \label{fig:subsets}
\end{figure*}

Figure \ref{fig:subsets} shows our results. The histograms show a notable difference in the distribution of bulge sizes for different mass and color subsets. Lower mass blue spirals have smaller bulges, while red and/or high mass spirals have larger bulges, especially massive red spirals. All subsets exhibit a similar skew in the distribution of arm winding scores towards tightly wound; massive spirals are slightly more tightly wound than lower mass spirals.

The trend between spiral arm winding and bulge size is shown for all subsets in the middle of Figure \ref{fig:subsets}. We observe no strong correlation in any sub-set. In low mass ($\log(M_\star/M_\odot< 10.3$), blue spirals (top panel of Figure \ref{fig:subsets}), 
we observe the positive correlation ($p=0.092$) noted by Hubble \citep{Hubble1926}. For the $\sim$\%20 of spiral galaxies that are red and/or massive ($\log(M_\star/M_\odot)>10.3$), the correlation  becomes negative ($p$=values from -0.045 to -0.174) -- looser arms appear mildly associated with larger bulges. This is most significant in the subset of massive red spirals (lower panel of \ref{fig:subsets}), which have the steepest negative slope. 

\section{Discussion and Conclusion}
We find that the observed correlation between arm windiness and bulge size in spiral galaxies has a significant mass and color dependence. Only low mass blue spirals show clear evidence for a Hubble sequence like correlation between bulge size and spiral pitch angle. 
 
The photographic plate images Hubble used would be more sensitive to bluer wavelengths and typical samples of very nearby galaxies skew towards lower mass, while SDSS contains both redder and more massive spirals.

The superposition of both low-mass blue and redder/more massive spirals perhaps explains the lack of correlation noted by \citet{Masters2019}. It is clear that sample selection (along with methods of determining pitch angle) may play an important role in what correlation is observed.\\

Our findings suggest that spiral galaxies with different masses and colors, properties which correlate with varying disc--halo mass fraction, or local environment (\citealt{Bamford2009}), may have different mechanisms of spiral arm formation. Spiral arm formation mechanisms can invoke tidal triggering, links to galactic bars, or self-excitation linked to shear \citep[][]{SellwoodMasters2022}. Redder massive spirals are more likely to be found in higher densities \citep[e.g.][]{Masters2010}, and show a greater diversity of rotation curves shapes (i.e. shear; \citealt{Frosst2022}). Treating all spiral galaxies as having the same spiral formation mechanism may obscure the physical process.

\bibliography{references}{}
\bibliographystyle{aasjournal}



\end{document}